\documentstyle[12pt]{article}
\begin{document}
\setcounter{footnote}{0}
\renewcommand{\thefootnote}{\fnsymbol{footnote}}
\begin{flushright}CTP-TAMU-2/97\\
NUB-TH-3152/97
\end{flushright}

\begin{center}
{\bf Detecting Physics At The Post-GUT And String Scales\\ 
By Linear Colliders}
~\\
~\\
~\\
R. Arnowitt\\Center for Theoretical Physics, Department of Physics\\
Texas A\& M University, College Station, TX  77843-4242\\
~\\
~\\
Pran Nath\\Department of Physics, Northeastern University\\
Boston, MA  02115\\
\end{center}
~\\
~\\
~\\
\begin{abstract}
The ability of linear colliders to test physics at the post-GUT scale is
investigated.  Using current estimates of measurements available at such
accelerators, it is seen that soft breaking masses can be measured with errors
of about (1-20)\%.  Three classes of models in the post-GUT region are
examined:  models with universal soft breaking masses at the string scale,
models with horizontal symmetry, and string models with Calabi-Yau
compactifications.  In each case, linear colliders would be able to test
directly theoretical assumptions made at energies beyond the GUT scale to a
good accuracy, distinguish between different models, and measure parameters that
are expected to be predictions of string models.
\end{abstract}
~\\
~\\
\noindent
{\bf 1.~~Introduction}
~\\

Much of current high energy theory considerations have centered
around the possibility that the aspects of the Standard Model
(SM) that are not presently understood (e.g. Yukawa couplings,
CKM parameters etc.) are consequences of new physical principles
arising at or near the Planck scale ($M_{P\ell} =
(\hbar c/8\pi G_N)^{1/2}\cong 2.4 \times 10^{18}$ GeV).  It is thus
important to ask whether hypotheses made at such high energies can be
experimentally verified.

One normally thinks that a high energy accelerator allows one to
learn about physics below its energy reach, but physics above this
energy is unprobed.  Thus, for example, LEP1 has determined the lower bound on
the Higgs mass to be 65 GeV, and one will have to wait until LEP2 to learn
more.  However, at least two theoretical results have moderated this
viewpoint.  First the renormalization group equations (RGE) allow one to take
data at one energy and extropolate it to a higher energy to test theoretical
ideas at this higher energy.  A second, related, result is supersymmetric
grand unification.  It implies that this upward extropolation can
be made over an enormous energy domain, i.e. of over 10$^{14}$ GeV.

The fact that the three coupling constants $\alpha_1, \alpha_2$ and
$\alpha_3$ unify for supersymmetric (SUSY) models to a GUT value
$\alpha_G\cong 1/24$ at a scale M$_G$ of about 10$^{16}$ GeV was
seen in the 1990 LEP precision data [1], appears quite firm.  Thus there have
been refinements in the data and refinements in the theoretical treatment (i.e.
inclusion of SUSY threshold effects at M$_S\simeq$ 100 GeV - 1 TeV [2], GUT
scale threshold effects [3,4], and small Planck scale effects [4]).  It is
possible that the unification of the three coupling constants at
M$_G\cong 2\times 10^{16}$ GeV is purely a numerical accident,
particularly since there is an adjustable parameter, the SUSY mass
scale, M$_S$.  However, such an accident is not too easy to
achieve for several reasons:  there is only a narrow window of values
for M$_G$ between about 5$\times 10^{15}$ GeV (below which in most
models a too rapid proton decay (p$\rightarrow e^+ \pi^0$) would occur)
and the ``string" scale about $5\times 10^{17}$ GeV, (above which gravitational
effects become strong invalidating the analysis.)  Further, unification does not
occur for the Standard Model, and satisfactory SUSY unification
occurs only for two Higgs doublets (the minimal number) and for no
more than four families.  Finally, naturalness requires
that $M_S\stackrel{<}{\sim}$ 1 TeV, which indeed turns out to be the
case.

In order to see whether grand unification is an accident or has deeper
physical significance, it is necessary to have other measurements to
experimentally test the idea.  We will see below that linear colliders
(LC) can indeed do this, and can not only probe GUT scale physics, but
also physics that may be occurring above the GUT scale, and do this with
a high degree of precision.

To analyse grand unification we use here the supergravity models [5] where
supersymmetry is broken in a hidden sector at a scale $\stackrel{>}{\sim}$
M$_G$.  These models have a number of positive attributes including
the following:  (1)  They account for the unification of the coupling
constants.  (2)  They allow for spontaneous breaking of supersymmetry at the
GUT (or Planck) scale.  (This is a crucial feature for without SUSY
breaking it is not possible to confront a theoretical model with
experiment.)  (3)  Using the RGE to go to low energies, one finds that
the spontaneous breaking of supersymmetry at energy $\stackrel{>}{\sim}$ M$_G$
triggers the spontaneous breaking of SU(2) x U(1) at the electroweak scale
M$_{EW}$ = O(M$_Z$) [6].  Thus supergravity models give an explanation of
the Higgs phenomena (and predicted that the top quark would be heavy i.e. 90 GeV$\stackrel{<}{\sim}
m_t\stackrel{<}{\sim}$ 200 GeV) a decade before its discovery.  The above three
items together imply a very predictive theory, and most of the phenomenological
SUSY analyses make use of some or all of the constraints implied by
supergravity.

In order to understand the nature of quantities in GUT models that might
be measured, we briefly summarize some of the structure of the
supergravity models.  These models depend on three functions of the
scalar fields $\phi_i(x)$ (representing sleptons, squarks, etc.):  the
gauge kinetic function f$_{\alpha\beta}(\phi_i$) (which enters in the
Lagrangian as f$_{\alpha\beta}F_{mu\nu}^{\alpha}F^{\mu\nu\beta}$ with
$\alpha,\beta$ = gauge indices), the Kahler
potential K($\phi_i,\phi_i^{\dagger}$) (which appears in the scalar kinetic
energy as $K^i_j\partial_{\mu}\phi_i\partial^{\mu}\phi_j^{\dagger}$,
K$^i_j\equiv\partial^2K/\partial\phi_i\partial\phi_j^{\dagger}$ and
elsewhere) and the superpotential W($\phi_i$).   The latter two enter only in
the combination

\begin{equation}
G~(\phi_i,\phi_i^{\dagger}) = \kappa^2 K~(\phi_i,\phi_i^{\dagger}) + \ell
n~[\kappa^6~\vert W(\phi_i)\vert^2]
\end{equation}
\noindent
where $\kappa= 1/M_{P\ell}$.  Writing
$\{\phi_i\}=\{\phi_{a}, z\}$ where $\{\phi_a\}$ are the physical sector fields
(quarks, leptons, Higgs) and $z$ are the superHiggs fields whose VEVs,
$\langle z\rangle = O~(M_{P\ell})$, break supersymmetry, the
superpotential decomposes into a physical and a hidden part

\begin{equation}
W~(\phi_i)= W_{phys}(\phi_a) + W_{hid}(z)
\end{equation}
\noindent
with $\kappa^2\langle W_{hid}\rangle = O(M_S)$.

The quantities f$_{\alpha\beta}$, K and W,  at the level of supergravity
theory, and are determined by a new physical principle that operates at the
Planck scale.  However, if one expands these
functions in a polynomial in $\phi_a$, those terms carrying the mass dimensions
of f$_{\alpha\beta}$, K and W are accessible to low energy discovery (their
coupling constants are dimensionless) while higher terms, representing Planck
physics corrections, are scaled by $\kappa$.  Thus 

\begin{equation}
f_{\alpha\beta}(\phi_i) = c_{\alpha\beta}(x) + \kappa c^a_{\alpha\beta}(x)
\phi_a + \frac{1}{2}\kappa^2 c^{ab}_{\alpha\beta}(x)\phi_a\phi_b+\cdots
\end{equation}

\begin{eqnarray}
K(\phi_i,\phi_i^{\dagger})&=&\kappa^{-2}c(x,y) +
c^a_b(x,y)\phi_a\phi_b^{\dagger}\nonumber\\
&+&~(c^{ab}~(x,y)~\phi_a\phi_b+h.c.)\nonumber\\
&+& (c^a_{bc}
(x,y)\phi_a\phi_b^{\dagger}\phi_c^{\dagger}+h.c.) +\cdots
\end{eqnarray}

\begin{equation}
W_{phys}(\phi_i) = \frac{1}{6}\lambda^{abc}
(x)\phi_a\phi_b\phi_c+\frac{1}{24}\kappa\lambda^{abcd}(x)\phi_a\phi_b\phi_c\phi_d
+\cdots
\end{equation}
\noindent
where x$\equiv\kappa$z, y$\equiv\kappa$z$^{\dagger}$, and the condition
K$^{\dagger}$ = K implies

\begin{equation}
c^a_b(x,y) = c^b_a(y,x)^{\dagger}; c (x,y) = c(y,x)^{\dagger}
\end{equation}

\noindent
The coefficients in the above expansions, c$_{\alpha\beta}$(x),
c$_b^a$(x,y), etc. have been scaled so that when super Higgs VEVs are taken
($<$x$>$, $<$y$>$ = O(1)) they are of O(1).  Also, as is well known [7], one may
always transfer the holomorphic c$^{ab}$ terms of (4) into the superpotential
by a Kahler transformation

\begin{equation}
W\rightarrow W' = W exp [\kappa^2c^{ab}\phi_a\phi_b] = W+\kappa^2 W
c^{ab}\phi_a\phi_b +\cdots 
\end{equation}

\noindent which gives rise to an effective $\mu$ term of the correct
electroweak size

\begin{equation}
\mu^{ab}(x) \phi_a\phi_b;~~~~~ \mu^{ab}(x) = \kappa^2 W_{hid} c^{ab}(x)
\end{equation}
\noindent
Thus $\mu^{ab}\equiv \langle \mu^{ab}(x) \rangle= O~(M_S)$, and the $\mu$
term arises naturally.  One may rescale
the gauge and chiral fields so that their kinetic energies have canonical
form, after which (for a simple gauge group)

\begin{equation}
\langle c_{\alpha\beta}(x) \rangle = \delta_{\alpha\beta};~~~~~~ \langle c^a_b
(x,y) \rangle = \delta^a_b;~~~~~~ \langle c_{xy} \rangle = 1
\end{equation}
\noindent
(where c$_{xy} = \partial^2 c/(\partial x\partial y)$.

The higher terms in Eqs. (3-5) scaled by $\kappa = 1/M_{P\ell},$ give rise
to non-renormalizable operators (NROs), emphasizing the fact that
supergravity models are effective field theories below the Planck scale. 
Indeed, it would be surprising if such terms did not exist (e.g. in
string theory one expects such NROs to arise upon integrating out the
tower of Planck mass states).  The non-zero gaugino masses at M$_G$

\begin{equation}
(m_{1/2)\alpha\beta}=\frac{1}{4}\kappa^{-3}~\langle e^{G/2}~
G^i (K^{-1})^j_i~f^{\dagger}_{\alpha\beta j}~\rangle
\end{equation}

\noindent
imply such structures occur in the expansion of c$_{\alpha\beta}$(x),
i.e. c$_{\alpha\beta}$ (x) = $\delta_{\alpha\beta} + \kappa
c^{(1)}_{\alpha\beta} z+\cdots$.  A non-zero c$_{\alpha\beta}^{(1)}$ is
required if $m_{1/2}$ is to be of O(M$_S$).  Thus it would not be
surprising if the corresponding term in the physical sector, $\kappa
c^a_{\alpha\beta}\phi_a$ were also present.  Such a term would not be
negligible if $\phi_a$ were the field that breaks the GUT group to the SM
group (e.g. the ${\underline{24}}$ of SU(5) [8]) for then $\langle\phi_a \rangle
= O(M_G)$ and this term is O (M$_G/M_{P\ell}$) i.e. a (1-10)\% correction to
the leading term.  Indeed, the current value, $\alpha_3~(M_Z)$ = 0.118 $\pm$
0.003 [9], suggest the existance of a few percent correction of this type
[4].  Thus the effects of Planck scale physics on the low energy domain may
have already been seen.

In this paper, we consider the possibility of using linear colliders to
investigate the post GUT regime.  In Sec. 2 we first review what may be
determined about GUT scale physics at colliders.  We then consider three
possible scenarios of the nature of post GUT physics.  In Sec. 3, we
examine the possibility that the soft breaking masses at the string scale
are universal (the RGE producing non-universal effects at M$_G$).  It is
shown there that a LC can distinguish between different gauge groups and
even determine the value of M$_{str}$. In Sec. 4 we examine a simple
horizontal group which determines the nature of non-universal soft
breaking.  In Sec. 5 we examine the Kahler potential arising in a class
of Calabi-Yau string models.  Sec. 6 contains conclusions.
~\\
~\\
\noindent
{\bf 2.~~GUT Scale Physics}
~\\

Aside from small GUT scale threshold corrections, most physics below
M$_G$ is insensitive to the nature of the physical sector GUT group $G$, if
$G$ breaks to the SM group at M$_G$.  Thus relatively model independent
tests of some GUT scale physics can be done.  

The gaugino soft breaking mass of Eq. (10) is

\begin{equation}
{(m_{1/2})}_{\alpha\beta} = \frac{1}{4}\kappa^{-1} \langle G^i
(K^{-1})^j_i f^{\dagger}_{\alpha\beta j} \rangle m_{3/2}
\end{equation}
\noindent
where m$_{3/2} = \kappa^{-1} \langle$ exp [G/2] $\rangle$ is the
gravitino mass.  In all models where the physical gauge group is a simple group
and the hidden sector fields are $G$ singlets, the gaugino masses will be
universal at M$_G$ to a very good approximation.  For this situation the
leading term occurs when i, j are both in the hidden sector yielding
${(m_{1/2})}_{\alpha\beta} = m_{1/2}\delta_{\alpha\beta}$
with 

\begin{equation}
m_{1/2} = \frac{1}{4} \langle [c_x(x) +
\kappa^{-1}\partial_xW_{hid}/W_{hid}]f_x \rangle m_{3/2}
\end{equation}

\noindent
where c$_{\alpha\beta}$(x)$\equiv\delta_{\alpha\beta}$ f (x).  Eq. (12)
implies $m_{1/2}$ = O (M$_S$), though m$_{1/2}$ could deviate
considerably from m$_{3/2}$.  Leading corrections to Eq. (12)
arise when j = a and i is in the hidden sector, giving from Eqs. (3,4) terms
proportional to $\kappa c^a_{by}c^{a\dagger}_{\alpha\beta}\phi^{\dagger}_b$.
If $\phi_b$ is a physical sector field whose VEV breaks G and hence is of
O(M$_G$), then one might expect a non-universal correction of order
M$_G/M_{P\ell}\approx$ (1-10)\%.  Note that this type of Planck scale term is
precisely the one entering the grand unification discussion above which can
lead to a reduction in predicted value of $\alpha_3(M_Z)$ [4].

Universality at $M_G$ leads to the well known relations for the U(1), SU(2)
and SU(3) gaugino masses at the electroweak scale [10]

\begin{equation}
{\tilde m}_i = (\alpha_i/\alpha_G)m_{1/2};~~~~ i = 1,2,3
\end{equation}
\noindent
with the gluino mass $m_{\tilde g}\cong {\tilde m}_3$ [11].  The
electroweak sector of this prediction can be well tested at the NLC, i.e.
to about 5\% [12].  Current analysis shows that m$_{\tilde g}$ can be
measured to within about (1-10)\% at the LHC (depending on the parameter
point) [13], allowing a good test of the third relation ${\tilde m}_2/{\tilde
m}_3 = \alpha_2/\alpha_3$ as well, and an experimental determination of the
GUT scale parameter m$_{1/2}$.  With improved precision one may even
be able to detect the small deviations of ${\tilde m}_i$ from universality and
correlate them with their effects on the grand unification predictions of
$\alpha_3(M_Z)$ mentioned above.

A second important prediction of these supergravity models is the
determination of $\mu^2$ at the electroweak scale from the RGE's:

\begin{equation}
\mu^2={m_{H_1}^2-m_{H_2}^2 tan^2\beta\over tan^2\beta-1} -\frac{1}{2}M_Z^2
\end{equation}

\noindent
where $m_{H_{1,2}}$ are the running Higgs masses at the electroweak scale
(including loop corrections) and tan $\beta = \langle H_2 \rangle/\langle
H_1\rangle$.  The $m_{H_{1,2}}^2(M_Z)$ are scaled by their values
$m_{H_{1,2}}(0)$ at M$_G$ and by m$_{\tilde g}$.  We restrict these to be
within the ranges 0 $\leq m_{H_{1,2}}(0)\leq$ 1 TeV, 180 GeV $<$ m$_{\tilde
g}~\leq$ 1 TeV, where the upper bound is a naturalness requirement, and
the lower bound on m$_{\tilde g}$ is the current experimental limit [12]. 
Hence over most of the parameter space $\mu$ is large and
$\mu^2/M_Z^2~>>$ 1 (i.e. $\mu\stackrel{>}{\sim}~3M_Z$).  In this domain, a set
of scaling laws hold for the neutralinos ($\tilde\chi_i^0$, $i = 1\cdots
4$), charginos ($\tilde\chi_i^\pm$, $i$ = 1,2) and Higgs bosons (h,
H$^0$, H$^{\pm}$, A) [14]:

\begin{equation}
2m_{\tilde\chi^0_1}\cong m_{\tilde\chi^0_2}\cong m_{\tilde
\chi^{\pm}_1}\cong~\biggl (\frac{1}{3}-\frac{1}{4}\biggr )~m_{\tilde g}
\end{equation} 

\begin{equation}
m_{\tilde\chi^0_3}\cong m_{\tilde\chi^0_4}\cong m_{\tilde
\chi^{\pm}_2}>>m_{\tilde\chi^0_1}
\end{equation}

\begin{equation}
m_{H^0}\cong m_H^{\pm}\cong m_A >>m_h
\end{equation}

\noindent
At a linear collider (LC) the $\tilde\chi^0, \tilde\chi^{\pm}$ masses can
be measured to $\sim$ 1\% [12,15] and probably good measurements of the Higgs
masses will also be available.  Thus if these relations were satisfied,
it would be good circumstantial evidence for the minimal supergravity
unification (though Eqs. (15-17) are not absolutely required by such models as
Eq. (14) also has solutions where $\mu$ = O(M$_Z$)).  It should be noted
that many of the particles in Eqs. (15-17) may lie beyond the reach of
the NLC, and so higher energy linear colliders will play an important role in
testing such relations.
~\\
~\\
\noindent
{\bf 3.~~Testing String Scale Universality}
~\\

At scales beyond M$_G$, the theory becomes sensitive to both the nature of the
GUT group and the particle spectrum above M$_G$ (which may no longer be just
that of the MSSM, i.e. only what was needed to achieve grand unification at
M$_G$).  Linear colliders will be sensitive to both these types of model
dependences, and hence help distinguish between different possibilities.  While
universality of the gaugino masses is an expected consequence of supergravity
grand unification with supersymmetry breaking at the Planck scale, the same is
not true for the other soft breaking masses.  These arise from the effective
potential which has the form [5]

\begin{equation}
V=e^{\kappa^2K}[(K^{-1})_i^j~(W^i+\kappa^2K^iW)~(W^j+\kappa^2K^jW)^{\dagger}-3\kappa^2\vert~W\vert^2]+V_D
\end{equation}
\noindent
where
\begin{equation}
V_D = \frac{1}{2} g_{\alpha}g_{\beta}
(Ref^{-1})_{\alpha\beta}~(K^i~(T^{\alpha})_{ij}~\phi_j)~(K^k~(T^{\beta})_{k\ell}~\phi_{\ell})^{\dagger}
\end{equation}
\noindent
where W$^i\equiv\partial W/\partial\phi_i$, T$^{\alpha}$ are the group
generators, $K^i=\partial K/\partial\phi_{i}$, (K$^{-1})^j_i$ is the inverse of
the Kahler metric, and g$_{\alpha}$ are coupling constants.  The scalar soft
breaking masses m$_0$ and the cubic A$_0$ and quadratic B$_0$ soft breaking
masses arise from two types of terms:  the superpotential in Eq. (18) which
does indeed yield a universal contribution, and the Kahler potential which may
or may not give a universal contribution [5,7,16].  In terms of the expansions
of Eqs. (3-5), one finds [5,7,16]
\begin{eqnarray}
(m_0^2)_b^a&=&\langle\delta_b^a~[\vert c_x \vert^2-2 + (c_x+c_x)
~(\kappa^{-1}~W_{hid}^x/W_{hid})+\vert\kappa^{-1}~W_{hid}^x/W_{hid}\vert^2]\nonumber\\
&+& [c_{cx}^ac_{by}^c - c_{bxy}^a]~[\vert
c_x\vert^2+(c_x+c_x^{\dagger})~(\kappa^{-1}~W_{hid}^x/W_{hid})+\nonumber\\
&+&\vert\kappa^{-1}~W_{hid}^x/W_{hid}\vert^2]\rangle~m^2_{3/2}
\end{eqnarray}
\noindent 
where W$_{hid}^x = \partial W_{hid}/\partial x$, and for simplicity
we have assumed only one field z grows a Planck mass VEV to break
supersymmetry:  $<$x$>$ = $\kappa^{-1}<$z$>$.  (One may easily generalize
this.)  Imposing the condition that the cosmological constant vanish, i.e.
$\langle$V$\rangle$ = 0, reduces m$_0^2$ to the simpler form 
\begin{equation}
(m_0^2)_b^a = [\delta_b^a+3\langle c_{cx}^a~c_{by}^c -
c_{bxy}^a\rangle ] m_{3/2}^2  
\end{equation} 
\noindent  
We see that m$_0$ is scaled by m$_{3/2}$ but can differ considerably from it.

Non-universal scalar masses can arise from c$_{cx}^a$ etc.,
i.e. from derivatives of the Kahler metric K$_b^a$ with respect to the
super Higgs fields z$\equiv\kappa x$ and z$^{\dagger}\equiv\kappa y$
[16].  These terms will be universal only if K$_b^a = \delta_b^a$K (x,y)
i.e. the super Higgs couples universally in K to the physical particles. 
One possibility is that the symmetry of the Kahler potential, which
controls the amount of universality, originates at the higher mass scale
where supersymmetry is broken.  The highest scale that one can still
treat this phenomena field theoretically is the string scale,
M$_{str}\simeq 5\times 10^{17}$ GeV.  In this section we consider then the
suggestion that K$_b^a=\delta_b^a$K at M$_{str}$ and the scalar masses
are universal at $\mu = M_{str}$ [17] (a possibility that can actually
occur in Calabi-Yau compactification of four dimensional superstrings
[18]).  The RGE from M$_{str}$ to M$_G$ would then lead to non-universal
soft breaking contributions, even if universality held at M$_{str}$. 
Such phenomena can then be tested at a LC since they produce effects at
low energy.  It is thus possible to explore experimentally the physics
between M$_G$ and M$_{str}$.  To illustrate this, we consider several
examples of GUT theories.
~\\
~\\
\noindent
(i) SU(5) GUT
~\\
~\\ 
\indent
We assume here for simplicity the minimal particle content above M$_G$, 
i.e. that matter exists in three generations of 10 = M$_i^{XY}$ and
${\bar 5}\equiv {\bar M}_{ix}$ representations (i = 1,2,3), and there
is a ${\bar 5}={\cal H}_{1X}$ and 5 = ${\cal H}_2^X$ of Higgs (which
contain the two light Higgs doublets coupling to matter) and a 24 =
$\sum_Y^X$ to break SU(5) to the SM.  (X,Y = 1$\cdots5$ are SU(5)
indices.)  The superpotential has the form (retaining only the large third
generation Yukawas),
\begin{eqnarray}
W &=& \biggl[\frac{1}{4} h_t\epsilon_{XYZWU}~M^{XY}~M^{ZW}~{\cal
H}_2^U+h_bM^{XY}{\bar M}_X{\cal H}_{1Y}\biggr]\nonumber\\
&+&\biggl[Mtr\Sigma^2+\frac{1}{6}\lambda_1tr\Sigma^3+\lambda_2{\cal
H}_1\Sigma{\cal H}_2 +\mu{\cal H}_1{\cal H}_2\biggr]
\end{eqnarray}
\noindent
If one were to assume that the soft breaking masses were universal at
M$_{str}$, then SU(5) invariance implies that there would be four soft
breaking masses at M$_G$ (which would then modify low energy phenomena). 
These are m$_{10}$ (which contains q$\equiv$ (${\tilde u} _L,{\tilde
d}_L$), u$\equiv{\tilde u}_R,~e\equiv{\tilde e}_R$), m$_5$ (which contains
$\ell\equiv ({\tilde\nu}_L, {\tilde e}_L$), d$\equiv{\tilde d}_R$) and
m$_{{\cal H}_{1,2}}=m_{H_{1,2}}$ where H$_{1,2}$ are the two light Higgs
doublets.  One may choose one of the soft breaking
masses as the reference mass, ${\tilde m_0}$, and consider deviations of
the other masses from ${\tilde m_0}$.  A convenient choice is
m$_{10}\equiv{\tilde m}_0$ and we write
\begin{equation}
m_5^2 = {\tilde m}_0^2~(1+\delta_5);~~m_{H_{1,2}}^2={\tilde
m}_0^2~(1+\delta_{1,2})
\end{equation}
\noindent
Using Eq. (22), one may calculate the expected deviations
from universality that result at M$_G$.  These are in general
significant ($\approx$ 50\%) though not enormous, and are sensitive to
the parameters of the model.
 
One can determine the values of $\tilde m_0$ and $\delta_5$ at a linear
collider and test the breakdown of universality experimentally
that occurs in the post-GUT regime.  Thus using the RGE to take the
masses down to the electroweak scale, one finds [19]
\begin{equation}
{\tilde m}_0^2 = m_{\tilde e_{R}}^2 - 0.151 m_{1/2}^2 + sin
^2\theta_WM_Z^2 cos 2\beta
\end{equation}
\begin{equation}
{\tilde m}_0^2\delta_5  = m_{{\tilde e}_L}^2-m_{{\tilde e}_R}^2 - 0.377 m_{1/2}^2 +
\biggl(\frac{1}{2}- sin^2\theta_W\biggr) M_Z^2 cos 2\beta
\end{equation}
\noindent
with $m_{1/2} = (\alpha_G/\alpha_2)~{\tilde m}_2$.  At the NLC
one expects to be able to measure $m_{{\tilde e}_{R}}$, $m_{{\tilde
e}_{L}}$ to about 1\%, ${\tilde m}_2$ to about 3\%, tan$\beta$ to 10\%
[12,15,20] and $\alpha_G$ to perhaps 3\%.  As an example, for
the case m$_{{\tilde e}_{L}}$ = 240 GeV, m$_{{\tilde e}_{R}}$ =
200 GeV, ${\tilde m}_2$ = 120 GeV and tan$\beta$ = 5, one finds from Eqs. (24,
25) and the estimated errors that  
\begin{equation}
{\tilde m}_0\cong (187\pm3)~GeV;~~~~~~\delta_5\cong 0.206\pm 0.031
\end{equation}
\noindent
Eq. (26) gives an indication of the remarkable level of accuracy
obtainable at a LC for the post-GUT parameters.  Further, there are
many other relations that can be used to determine ${\tilde m}_0$ and
$\delta_5$.  For example, one can use squarks instead of sleptons since
their masses can also be measured at the NLC to 1\% provided ${\tilde
m_{\tilde q}}< m_{\tilde g}$ [21] and $m_{\tilde q}$ is within the reach
of the NLC.  The difference $m_{\tilde u_{L}}^2 - m_{\tilde d_{L}}^2$
determines $\delta _5$ to a similar accuracy.  The many different ways
to determining $\tilde m_0$ and $\delta_5$ would cross check the
validity of the SU(5) model.

The parameters $\delta_1$ and $\delta_2$ enter sensitively into $\mu$
and m$_A$.  In the scaling region of Eqs. (7-9), $\mu$ can be
accurately determined from the $\tilde\chi_2^{\pm}$ and
$\tilde\chi_{3,4}^0$ masses provided the LC has high enough energy to
reach these thresholds [21].  Measurement of m$_A$ requires that the A
be pair produced at the LC [22].  The relations one could use to
determine $\delta_1$ and $\delta_2$ are
\begin{eqnarray}
\mu^2 (t^2-1)&=&\biggl[\delta_1-\frac{1}{2}t^2(1+D_0)\delta_2\biggr]{\tilde
m}_0^2+\biggl[ 1-\frac{1}{2} t^2(3D_0-1) \biggr] \tilde m_0^2\nonumber\\
&+&[0.528 + t^2 (3.22 - 3.80 D_0 + 0.060 D_0^2]~m_{1/2}^2\nonumber\\
&+&\frac{1}{2} t^2 (1-D_0) \frac {A_R^2}{D_0} - \frac{1}{2} M_Z^2 (t^2
- 1)
\end{eqnarray}
\begin{eqnarray}
m_A^2\biggl(\frac{t^2-1}{t^2+1}\biggr) &=& [\delta_1-\frac{1}{2}
(1+D_0)~\delta_2 + \frac{3}{2}~(1-D_0)\biggr]~\tilde m_0^2\nonumber\\
&+& [3.22-3.80 D_0+0.060D_0^2]~m_{1/2}^2 + \frac{1}{2}
(1-D_0)\frac{A_R^2}{D_0}\nonumber\\
&-& \frac{t^2-1}{t^2+1}~M_Z^2
\end{eqnarray}

\noindent
where t$\equiv$ tan$\beta$, D$_0$ is the Landau pole denominator, $D_0 =
1-m_t^2/m_f^2$, $m_f\cong~200 sin\beta$ GeV, and A$_R$ is the residue at
the pole, $A_R\cong -A_t - 1.74~m_{1/2}$.  (The numerical coefficients
come from running the RGE from M$_G$ to the electroweak scale.)  Thus in
order to determine $\delta_1$ and $\delta_2$, one needs to know $A_t$, and
this could be determined from the light stop ($\tilde t_1$)production
cross section [13,23].  As an example we consider the parameters m$_0$
= 200 GeV, $\tilde m_2$ = 120 GeV ($\tilde
m_2\equiv~(\alpha_2/\alpha_G) m_{1/2})$, $\mu$ = 325 GeV, m$_A$ = 400
GeV, A$_t$ = -- 0.5 m$_0$, tan$\beta$ = 5 and m$_t$ = 175 GeV.  We assume
m$_0$, m$_A$ and $\mu$ are determined with $\pm$ 2\% error, $\tilde
m_2$ with $\pm$ 3\% error, A$_R$ with $\pm$ 5\% error and tan$\beta$ with
$\pm$ 10\% error.  One finds 
\begin{equation}
m_{H1} (0) = (256\pm 15) GeV; ~~~~~~~ m_{H_{2}} (0) = (144\pm 35) GeV
\end{equation}
\noindent
which corresponds to $\delta_1$ = 0.634 $\pm$ 0.220 and $\delta_2$ =
0.485 $\pm$ 0.178.  Thus deviations from universality would be clearly
observable for this situation.

There are numerous other experimental tests one can put this SU(5) model
to:  (i) There are three mass differences where non-universal effects
cancel out:

\begin{equation}
m_{\tilde u_{L}}^2 - m_{\tilde u_{R}}^2, ~~m_{\tilde u_{L}}^2 - m_{\tilde
e_{R}}^2, ~~m_{\tilde e_{L}}^2 - m_{\tilde d_{R}}^2~~.
\end{equation}
\noindent
These quantities depend only on $\tilde m_0^2$ and the known RGE form factors in going from
M$_G$ to the electroweak scale.  (ii) One can examine generational
dependences in these relations and others discussed above.  (We have
suppressed generational indices in the above equations.)  In this way one can
check on the symmetry of the Kahler potential.  Thus there are many ways of
testing such an SU(5) model in the post-GUT regime.

We discuss now how one may determine M$_{str}$.  Assuming that the particle
spectrum is that of Eq. (22), one may use the RGE to run the soft SUSY
breaking masses to higher scales, and they would unify at $\mu = M_{str}$.  It
can then determine experimentally the value of M$_{str}$.  The
value of M$_{str}$  is one of the fundamental parameters of string
theory, and it is truly remarkable that it is accessible to experimental
test at linear colliders.  
~\\
~\\
\noindent
(ii)~~SO(10) GUT
~\\

In SO(10) models, each family is put into an SO(10) 16-plet representation,
which decomposes into its SU(5) content as 16 = 10 +${\bar 5}$ + 1, (the SU(5)
singlet being $\nu_R$).    There are a number of ways in which SO(10) can break
to the SM group.  We consider here the simplest variant where SO(10) breaks
directly to SU(3) x SU(2) x U(1) at M$_G$.  One might have expected that SO(10)
symmetry implies that m$_ 5$ = m$_{10}$ since the 10 and ${\bar 5}$ are part of
the same 16 representation.  However, SO(10) is a rank 5 group and the SM group
is rank 4.  When one breaks a higher rank group to a lower one, additional D
terms can effect the mass relations [24, 25].  Thus at M$_G$, one has
m$_{10}~\not= m_5$.  We will assume here again the simplest possibility that
the 5 and $\bar 5$ Higgs of SU(5) lie in the same 10 of SO(10) (i.e.
10 = 5 +$\bar 5$).  Then at M$_G$ one has [25]
\begin{equation} 
 m_{10}^2 = m_{16}^2 +\frac{1}{4} \biggl( m_{H_{1}}^2 -
m_{H_{2}}^2\biggr)
\end{equation}
\begin{equation}
m_5^2=m_{16}^2-\frac{3}{4} \biggl( m_{H_{1}}^2-{H_{2}}^2\biggr)
\end{equation}
\noindent
where, as in SU(5), $m_{10}=m_q=m_u=m_e$ and $m_5=m_{\ell}=m_d$. 
It is again convenient to chose our reference mass as $\tilde
m_0\equiv m_{10}$ with non-universal deviations parameterized as
$m_5^2 ={\tilde m}_0^2~(1+\delta_5)$, $m_{H_{1,2}}^2={\tilde
m}_0^2~(1+\delta_{1,2})$.  One has then that $m_{16}^2={\tilde
m}_0^2~[1-\frac{1}{4}~(\delta_2-\delta_1)]$ and 
\begin{equation}
\delta_5=\delta_2-\delta_1
\end{equation}
\noindent
In addition, the other SU(5) relations discussed above still hold. 
Eq. (33) represents the one additional constraint in SO(10) for
this pattern of symmetry breaking.  It would be testable to about 20\% at
a LC.  For example, using the numbers calculated in Eqs. (26, 29) one finds
$\delta_5/(\delta_2-\delta_1)$ = 0.160$\pm$ 25\%, showing for that case that
the SO(10) relation would be significantly violated. 
~\\ 
~\\
\noindent
(iii)~~SU(3) x SU(2) x U(1)
~\\

Some string models assume that the SM gauge group holds all the
way up to the string scale, after which unification will occur at
M$_{str}$ [26].  The RGE will then split the soft breaking masses
at M$_G$.  Choosing here $m_q\equiv{\tilde m_0}$ as the reference
mass one has

\begin{eqnarray}
m_u^2&=&\tilde m_0^2~(1+\delta_u);~~~m_e^2=\tilde
m_0^2~(1+\delta_2)\nonumber\\
m_d^2&=&\tilde m_0^2~(1+\delta_d);~~~m_{\ell}^2=\tilde
m_0^2~(1+\delta_{\ell})\nonumber\\
m_{H_{1,2}}^2&=&\tilde m_0^2~(1+\delta_{1,2})
\end{eqnarray}
\noindent
with the notation $u =\tilde u_R$, e = $\tilde e_R$ etc. as in
SU(5).  There are a priori no relations between the different
$\delta$'s and hence mass differences that were universal for
SU(5) or SO(10), i.e. Eq. (30), will no longer in general be
universal.  The different $\delta$'s can, of course, be measured at a
LC.  Thus $m_{\tilde u_{L}}^2-m_{\tilde u_{R}}^2$ will determine
$\delta_u$, $m_{\tilde e_{R}}^2$ determines $\delta_e$, etc.
~\\
~\\
\noindent
(iv)~~Distinguishing Post-Gut Groups
~\\

The LHC can give information about the nature of physics beyond the
GUT scale, and in fact distinguish between different gauge groups
that may hold beyond M$_G$.  In the examples discussed above, one
would conclude that if $\delta_{u,e}{\not=} 0$ or
$\delta_d{\not=}\delta_{\ell}$, then SU(5) and SO(10) would not be
valid GUT groups, but the SM group could still hold above M$_G$. 
But if, $\delta_u = 0 = \delta_e,~\delta_d = \delta_{\ell}$, then
both SU(5) and SO(10) would be consistent with this result.  But
if, $\delta_5{\not=}\delta_2-\delta_1$, then the specific SO(10) model
considered would be eliminated.  Similar considerations can be carried out for
other gauge groups and other symmetry breaking patterns.   

~\\
\noindent
{\bf 4.~~Horizontal Symmetry}
~\\
~\\
\indent
As discussed in Sec. 2, non-universality of soft breaking masses is controlled
by the structure of the Kahler potential.  We consider here a model where K
possesses a horizontal SU(2)$_H$ symmmetry [27] and is based on the total gauge
group

\begin{equation}
SU(5)\times SU(2)_H
\end{equation}
~\\
\noindent
While this model is not completely satisfactory phenomenologically, it does
illustrate which aspects of such ideas would be accessible to a LC.

We assume that the first two generations form an SU(2)$_H$ doublet, and the
third generation is a singlet.  The matter is in the usual SU(5) 10 and ${\bar
5}$ representations, while the Higgs are SU(2)$_H$ singlets in 5+${\bar 5}$ and
24 representations (where the 24 breaks SU(5) to the SM at M$_G$).  In addition
it is assumed that there are three SU(5) singlet, SU(2)$_H$ doublet Higgs,
$\phi^i_{(r)},$ r = 1,2,3, whose VEVs have the form
($\langle\phi_{(1)}\rangle,$ 0) and (0, $\langle\phi_{(2)}\rangle$) to break
the SU(2)$_H$ [27].  The superpotential then is 

\begin{eqnarray}
W&=&\biggl [\lambda_{ab}^1~M_a^{XY}{\bar M}_{bx}{\bar
H}_Y+\lambda^2_{ab}\epsilon_{XYZWU}M_a^{XY} M_b^{ZW}~H^U\biggr ]\nonumber\\
&+&\biggl [\lambda^3_r\kappa~\phi_{(r)}^iM_i^{XY} {\bar M}_X {\bar H}_Y
+\lambda^4_r\epsilon_{XYZWU} \kappa \phi_{(r)}^i M_i^{XY}M^{ZW}H^U+\cdots\biggr
]~~~~ 
\end{eqnarray}

\noindent
where a,b = 1,2,3, i,j = 1,2 are SU(2)$_H$ doublet generation indices, and
matter fields without generation subscripts are third generation SU(2)$_H$
singlets.  One may make a bi-unitary transformation to diagonalize
$\lambda^1_{ab}$ to the form diag $\lambda^1_{ab}$ = $\biggl(\lambda^d$, 
$\lambda^d$, $\lambda^b\biggr)$ while the anti-symmetry of the $\epsilon$-symbol
implies diag $\lambda^2_{ab}= (0,0,\lambda^t)$.  The second bracket in Eq. (36)
gives rise to mixing between the third and first two generations in the quark
mass matrix of size
\begin{equation}
\epsilon = O~(\kappa\langle\phi_{(r)}\rangle)
\end{equation}

\noindent
and thus second generation masses of size O($\epsilon^2$).  Hence reasonable
first and second generation quark masses require [27] $\lambda^d<< 1$ and
\begin{equation}
\epsilon\approx 1/10
\end{equation}
\noindent
Thus the picture this model presents is that supersymmetry breaks (in the
hidden sector) at the Planck scale ($\langle z\rangle\approx M_{P\ell}$),
SU(2)$_H$ breaks at the string scale ($\langle\phi_{(r)}\rangle\approx 1/10
M_{P\ell}$) and SU(5) at the GUT scale ($\langle\Sigma\rangle\approx
M_G\approx 1/100 M_{P\ell}$), and we will assume all this in the following.

The SU(2)$_H$ Higgs fields can produce corrections to the SU(5) gauge
function, but since the $\phi_{(r)}$ are SU(5) singlets, they have the form
\begin{equation}
f_{\alpha\beta}\left (\phi_{(r)}\right ) = \delta_{\alpha\beta}~\kappa^2
c_{rs}(x)\phi^i_{(r)} \epsilon
_{ij}\phi^j_{(3)}
\end{equation}
\noindent
These corrections are small, i.e. O ($\epsilon^2$), and maintain
the universality of the gaugino masses at the string scale, and down to
the GUT scale.  Thus they can be neglected.  The Kahler potential has the
expansion

\begin{eqnarray}
K&=&\kappa^{-2} c_0^0~(x,y) +[c_s^{10} M^{XY} M^{XY\dagger}+c_d^{10} M_i^{XY}
M_i^{XY\dagger}+c_s^{\bar 5} {\bar M_X}{\bar M_X^\dagger}\nonumber\\
&+&c_d^{\bar 5}{\bar M_X^i}M_X^{i\dagger}]+[c_H H^X H^{X\dagger}+c_{\bar H}{\bar
H_X}{\bar H_X^{\dagger}}+c_{\Sigma}\Sigma^X_Y\Sigma^{X\dagger}_Y]\nonumber\\
&+&[c^{10}_{(r)}\kappa\phi^i_{(r)} M_i^{XY}M^{XY\dagger}+c^{\bar
5}_{(r)}\kappa\phi^i_{(r)}{\bar M_{Xi}}{\bar M_X^{\dagger}}+h.c.]\nonumber\\
&+&\frac{1}{2}\biggl[ c^{10}_{(rs)}\kappa\phi^i_{(r)}
M_i^{XY}\kappa\phi_{j(rs)}M_j^{XY\dagger}+{\tilde
c}^{10}_{(r)}\kappa^2\phi^i_{(r)}\phi_{i(s)}M^{XY}M^{XY\dagger}\biggr]
\nonumber\\ 
&+&\cdots
\end{eqnarray}

\noindent
where the subscripts (s,d) stand for SU(2)$_H$ (singlet, doublet).  

The first two brackets in Eq. (40) are the SU(2)$_H$ invariant terms which
give rise to universal soft breaking masses for the first two generations in
the 10 and ${\bar 5}$ representations, but are, however, split from the singlet
third generation and from the Higgs soft breaking masses.  Thus in a notation
analogous to Eq. (23) they give rise to soft breaking masses

\begin{eqnarray}
(m_5^i)^2 &=& {\tilde m}_0^2 (1+\delta^d_5);~~~(m_5)^2 = {\tilde
m_0^2}~(1+\delta^s_5);\nonumber\\
m_{10}^2&=&{\tilde m}_0^2~(1+\delta_{10}^s);~m_{H_{1,2}}^2 = {\tilde m_0^2}
(1+\delta_{1,2})
\end{eqnarray}

~\\
\noindent
where the reference mass ${\tilde m_0}$ is now chosen to be the common mass of
the doublet of squarks in the 10 representation.  The third and fourth
brackets of Eq. (40) give rise to $\epsilon$ and $\epsilon^2$ breakings of
SU(2)$_H$ at M$_{str}$, the former arising only in the mixing of the third
(SU(2)$_H$ singlet) generation with the doublets.  Using Eq. (21), one has that
the mass matrix for example, for the d$_L$ squarks at M$_{str}$ is of the
form 
~\\
~\\

\begin{equation}
\tilde m_{d_{L}}^2=\left(\begin{array}{ccc}{\tilde
m_d^2}+\epsilon^2m^2_{11}&\epsilon^2m_{12}^2&\epsilon m_{13}^2\nonumber\\
\epsilon^2m_{12}^2&{\tilde m_d^2}+\epsilon^2m_{22}^2&\epsilon
m_{23}^2\nonumber\\ 
\epsilon m_{13}^2&\epsilon m_{23}^2&{\tilde
m_b^2}+\epsilon^2m_{33}^2
\end{array}\right)\end{equation}
~\\
~\\
\noindent
where $\tilde m_d^2=\tilde m_0^2~(1+\delta^d_5)$, $\tilde m_b^2=\tilde m_0^2
(1 +\delta_5^s$) and m$_{ij}$ = O(m$_{3/2}$).  Eq. (42) implies that the
first two generation masses at M$_{str}$ are split by only O($\epsilon^2$)
from their values of Eq. (41).  The smallness of the splitting, which is
natural for these models, is necessary to supress FCNC [27].

We now discuss what parts of the post GUT hypotheses of these models are
directly accessible to experimental test.  As seen in Eq. (26), $\tilde m_0$ is
determinable to perhaps 2\%, and so the O($\epsilon^2)\approx 10^{-2}$
splitting of the SU(2)$_H$ doublets would not be observable without a
significant improvement of measurement technique.  However, it should be
possible to distinguish this class of models from those of Sec. 3.  While
the non-universal effects in the (mass)$^2$ differences of Eq. (30) will still
cancel out if both masses are in the doublet (first two generation) or singlet
(third generation) SU(2)$_H$ representations, they will not cancel if one is a
doublet and the other is a singlet.  Thus, the first two differences of Eq.
(30) for this ``doublet-singlet" type difference determine $\delta^s_{10}$, and
if the model is correct, the value obtained should be the same to
O($\epsilon^2$) for each such difference. (There are eight independent
measurements of $\delta_{10}^s$ that should produce the same value of
$\delta^s_{10}$.)  The last difference, when one sfermion is in the doublet and
one in the singlet determines $\delta_5^d-\delta_5^s$ which one expects to be
non-zero (and there are four independent measurements which should give the
same value for this quantity).  As can be seen from Eq. (26), one expects the
values of $\delta_{10}^5$, $\delta_5^d$, $\delta_5^s$ to be determined to about
15 \% accuracy, which should allow good tests of the model.  Also, unlike the
models of Sec. 3, one does not expect the soft SUSY breaking masses to become
equal as one extropolates upwards towards M$_{str}$.  Thus while the very small
($\approx$ 1\%) effects of the breaking of SU(2)$_H$ are difficult to directly
measure, the general constraints of the SU(2)$_H$ symmetry of the Kahler
potential should be testable to a reasonable accuracy.  Physical assumptions
made in theories of this type at energies above M$_G$ (e.g. at
M$_{str}$) can be explicitly checked at a LC, and such models distinguished
from other models.
~\\
~\\
\noindent
{\bf 5.~~Superstring Models}
~\\
~\\
\indent 
The mechanism of supersymmetry breaking in superstring theory is not yet
understood, and as a consequence it is not possible to make phenomenological
predictions in string theory from first principles.  However, it has been
suggested that supersymmetry breaking may arise from dilaton(S) and
moduli (T$_i$,U$_i$) VEV formation.  With this assumption, it is possible
to calculate soft breaking parameters at the string scale in terms of
these unknown VEVs, and this has led to a large amount of analysis in the
literature.  (See e.g. [28,18].)  We consider in this section models
of this type arising in Calabi-Yau compactifications with (2,2) vacua
based on the gauge group E$_6\times$ E$_8$, matter then being in 27 and
$\overline{27}$ representations.  While the models considered here do not
lead to phenomenologically realistic predictions, they will allow us to
examine how accurately these string assumptions can be verified
experimentally by a LC.

For the case where there is only a single modulus T [29], or when there are
many moduli with equal $\mid F^{T_i}\mid$ terms [18], the soft breaking
masses are universal at the string scale and have the general form (for
vanishing U moduli F-terms) [18]:
\begin{equation}
m_{1/2}={\sqrt 3} sin\theta e^{-i\gamma_{S}}m_{3/2}
\end{equation}
\begin{equation}
m_0^2= [sin^2\theta + (cos^2\theta)~\Delta (T,T^*)]m^2_{3/2}
\end{equation}
\begin{equation}
A_0= -{\sqrt 3} [sin\theta e^{-i\gamma_{S}} +cos\theta
e^{-i\gamma_{T}}\omega (T,T^*)]m_{3/2}
\end{equation} 
\noindent
In Eqs. (43-45), the angle $\theta$ parameterizes the direction between the
Goldstino and the dilaton, $\Delta$ and $\omega$ include the $\sigma$-model
contribution and instanton correction to the Kahler potential, and
$\gamma_S, \gamma_T$ are possible CP violating phases.  In the following
we will for simplicity set, $\gamma_{S,T}$ to zero.  The quantities
$\theta, \Delta$ and $\omega$ are model dependent, and we will leave
them arbitrary for the moment.

The models considered here are examples of those of Sec. 3 with specific
string theory constraints.  As discussed in Sec. 2, m$_{1/2}$ can be
determined at a LC with error of about 5\%, and from Eq. (26), m$_0$ can be
determined with error of about 2\%.  Eqs. (43,44) imply
\begin{equation}
\frac{m_o^2}{m_{1/2}^2} = \frac{1}{3}~\left[1+\Delta ctn^2\theta\right]
\end{equation}

\noindent
and using the parameters of Sec. 2(ii) (${\tilde m_2}$ = 120 GeV, m$_0$ = 187
GeV) one finds

\begin{equation}
\Delta ctn^2\theta = 3.73\pm 0.25
\end{equation}
\noindent
Eqs. (43,45) give
\begin{equation}
\frac{A_0}{m_{1/2}} = -1-\omega ctn\theta
\end{equation}
\noindent
One may relate A$_0$ to A$_t$ by the RGE:
\begin{equation}
A_0 = \frac{A_R}{D_0} - 2.20 m_{1/2}
\end{equation}
\noindent
where A$_R$ = -A$_t$ -1.74 m$_{1/2}$ is the residue at the Landau pole,
and D$_0$ =1-$m_t^2/m_f^2$ where m$_f\cong$ 200 sin$\beta$ GeV [30]. 
For the parameter choice A$_t$ = -285 GeV and tan$\beta$ = 5 (with errors
of 5\% for A$_R$ and 10\% for tan$\beta$ as in Sec. 3) one finds
\begin{equation}
\frac{A_0}{m_{1/2}}= -1.539\pm 0.047
\end{equation}
\noindent
and hence from (48) one has
\begin{equation}
\omega ctn\theta =0.539\pm 0.047
\end{equation}

Specific Calabi-Yau compactifications determine the values of $\Delta$ and
$\omega$.  Eqs. (47) and (51) then allow for two independent experimental
determinations of ctn$\theta$ to check the validity of a given model.  We
consider two examples of models discussed in [18].
~\\
~\\
\noindent
(i)~~One-modulus Models

The values of $\Delta$ and $\omega$ for the four one-modulus models of [29]
can be calculated in the large Calabi-Yau radius limit [31].  In this limit,
the instanton contributions are negligible, and for ReT =5, $\Delta$ and
$\omega$ have average values of [18] $\Delta\cong$ 0.40, $\omega\cong 0.17$. 
Eqs. (47) and (51) then give respectively
\begin{equation} 
\mid ctn\theta\mid = 3.05\pm 0.14
\end{equation}
\begin{equation}
ctn\theta = 3.17 \pm 0.28
\end{equation}
\noindent
We see that for these parameters, the value of ctn$\theta$ is well
determined at a LC, and the two values are consistent with each other
(with the choice ctn $\theta >$ 0).  One may now return to (43) and
evaluate m$_{3/2}$.  Thus Eq. (52) yields 
\begin{equation}
sin\theta= 0.311\pm 0.013
\end{equation}
\noindent
and hence
\begin{equation}
m_{3/2}= (276\pm 18) GeV
\end{equation}
\noindent
Of course, other tests of the validity of these models can be made, such as
those discussed in Sec. 3 (where now the gauge group is E$_6$) and elsewhere
[28].  Both sin$\theta$ and m$_{3/2}$ are aspects of supersymmetry breaking. 
When a string understanding this phenomena becomes known, these
quantities would presumably be predicted by the model.  Thus Eqs. (54)
and (55) would then represent precision tests of the string picture of
SUSY breaking. 
~\\
~\\
\noindent
(ii)~~Maximum $\Delta$ Model
~\\
\indent
A model which maximizes the value of $\Delta$ occurs when ImT =1/4.  Then
for ReT =5 , Ref. [18] finds $\Delta$ = 1.62 and $\mid\omega\mid$ = 0.64. 
Eqs. (47) and (51) would now yield respectively
\begin{equation}
\mid ctn\theta \mid = 1.516\pm0.071;~~~\mid ctn\theta \mid= 0.842\pm 0.073
\end{equation}
~\\
\noindent
In this case the two determinations of ctn$\theta$ are
inconsistent, which would imply that this model is experimentally ruled out
for the given choice of low energy parameters.

We see from the above discussion that a LC is capable of testing the validity
of different string compactifications as well as being able to distinguish
among different compactifications.  Further, these determinations can be made
with very good accuracy.  In particular, one can test those assumptions that
are specifically string related.
~\\
~\\
\noindent
{\bf 6.~~Conclusions}
~\\
~\\ 
\indent
It is generally expected that the LHC and NLC will be able to unravel the
physics that lies above the Standard Model.  Thus if supersymmetry is correct,
these machines should be able to observe much of the SUSY mass spectrum, as
well as test GUT scale assumptions.  However, supergravity grand unification
is an incomplete theory, and many of the hypotheses used there presumably
reside in a more fundamental theory that exists above M$_G$.  A remarkable
feature of linear colliders is that they will be able to test theoretical
assumptions in this post-GUT domain.

In this paper, we have examined three classes of such post-GUT models: 
supergravity models with universal soft breaking at the string scale
M$_{str}$, models with SU(2)$_H$ horizontal symmetry, and Calabi-Yau string
models.  In each of these it was seen that theoretical assumptions made
at post-GUT scales could be checked with generally very good accuracy by
a LC.  Thus in the first class of models the predicted loss of
universality at M$_G$ could be well measured.  Different gauge groups,
e.g. SU(5), SO(10) could be distinguished, and the value of M$_{str}$
could be determined.  For the SU(2)$_H$ model, the very small splittings
resulting from the breaking of SU(2)$_H$ will probably require a
reduction by a factor of 5-10 in currently expected errors to be observed
at the LC.  However, the general SU(2)$_H$ symmetry should be
easily observable.  Finally, different Calabi-Yau compactifications would
be distinguishable, and in the cases considered, accurate direct
measurements of such explicitly string quantities as the partition of
the goldstino between the dilaton and the moduli, and the value of the
gravitino mass are obtainable.

However, the reach of the NLC with $\sqrt s$ = 500 GeV, while good for
studying light neutralinos, charginos and perhaps sleptons is likely to be
insufficient for the heavier SUSY particles needed to give a full knowledge of
what is happening at energies $\stackrel{>}{\sim}$ M$_G$.  For example, if LEP
1.9 does not discover the lightest chargino $\tilde\chi_1^{\pm}$, then
$m_{\tilde\chi_{1}^{\pm}}\stackrel{>}{\sim}$ 90 GeV.  In the scaling domain,
this would imply $m_{\tilde g}\stackrel {>}{\sim}$ 270 GeV.  The $\tilde u_L$
mass is given by $m_{\tilde u_{L}}^2\cong {\tilde m_0^2}$+0.893  
$m_{\tilde g}^2+\hfill\\  \left(\frac{1}{2}-\frac{2}{3}
sin^2\theta_W\right) M_Z^2 cos 2\beta$ which implies $m_{\tilde u_{L}}
>$ 250 GeV, with similar results for other squarks.  Thus in order to
sample the full SUSY spectrum, one needs colliders with $\sqrt s >$ 1
TeV (preferably up to $\sqrt s$ = 2 TeV). ~\\
~\\
\noindent
{\bf Acknowledgements}
~\\
\indent
This work was supported in part by NSF grants PHY-9411543 and
PHY-9602074.  A preliminary version of this work was presented at
``Physics With High Energy e$^+$ -- e$^-$ Colliders", Brookhaven National
Laboratory, May 1996 and appeared in ``Physics and Technology of the
NLC:  Snowmass 96", hep-ex/9605011.

~\\ 
\noindent  
{\bf References}
\begin{enumerate} 
\item 
P. Langacker, Proc. PASCOS90, Eds. P. Nath and S.
Reucroft (World Scientific, Singapore 1990); J. Ellis, S. Kelley and D.V.
Nanopoulos, Phys. Let.  {\bf B249}, 441 (1990), {\bf B260}, 131 (1991); U.
Amaldi, W. De Boer and H. Furstenau, Phys. Lett. {\bf B260}, 447 (1991); F.
Anselmo, L. Cifarelli, A. Peterman and A. Zichichi, Nuov. Cim. {\bf 104A}, 1817
(1991); {\bf 115A}, 581 (1992). 
\item 
J. Bagger, K. Matchev and D. Pierce,
Phys. Lett. {\bf B348}, 443 (1995); P.H. Chankowski, Z. Pluciennik, and S.
Pokorski, Nucl. Phys. {\bf B439}, 23 (1995). 
\item
R. Barbieri and L.J. Hall, Phys. Rev. Lett. {\bf 68}, 752 (1992); L.J. Hall
and U. Sarid, Phys. Rev. Lett. {\bf 70}, 2673 (1993); P. Langacker and N.
Polonsky, Phy. Rev. {\bf D47}, 4028 (1993).
\item
T. Dasgupta, P. Mameles and P. Nath, Phys. Rev. {\bf D52}, 5366 (1995); D.
Ring, S. Urano and R. Arnowitt, Phys. Rev. {\bf D52}, 6623 (1995); S. Urano,
D. Ring and R. Arnowitt, Phys. Rev. Lett. {\bf 76}, 3663 (1996); P. Nath, Phys.
Rev. Lett. {\bf 76}, 2218 (1996)
. \item
A.H. Chamseddine, R. Arnowitt and P. Nath, Phys. Rev. Lett. {\bf 49}, 970
(1982).  For reviews see P. Nath, R. Arnowitt and A.H. Chamseddine, ``Applied
N = 1 Supergravity" (World Scientific, Singapore 1984); H.P. Nilles, Phys.
Rep. {\bf 100}, 1 (1984); R. Arnowitt and P. Nath, Proc. VII Swieca Summer
School, ed. E. Eboli (World Scientific, Singapore 1994).
\item
K. Inoue et al., Prog. Theor. Phys. {\bf 68}, 927 (1982); L. Iba${\tilde n}$ez
and G.G. Ross, Phys. Lett. {\bf B110}, 227 (1982); L. Alvarez-Gaum\'e, J.
Polchinski and M.B. Wise, Nucl. Phys. {\bf B221}, 495 (1983); J. Ellis, J.
Hagelin, D.V. Nanopoulos and K. Tamvakis, Phys. Lett. {\bf B125}, 2275
(1983); L. E. Iba${\tilde n}$ez and C. Lopez, Phys. Lett. {\bf B128}, 54
(1983); Nucl. Phys. {\bf B233}, 545 (1984); L.E. Iba${\tilde n}$ez, C. Lopez
and C. Mu${\tilde n}$os, Nucl Phys. {\bf B256}, 218 (1985).
\item
S. Soni and A. Weldon, Phys. Lett. {\bf B126}, 215 (1983).
\item
C.T. Hill, Phys. Lett. {\bf B135}, 47 (1984); Q. Shafi and C. Wetterich,
Phys. Rev. Lett. {\bf 52}, 875 (1984).
\item
P. Langacker, talk at APS Division of Particles and Fields 1996 Conference
(DDF96 Minneapolis, August 1996).
\item
P. Nath, A.H. Chamseddine and R. Arnowitt, Proc. of 1983 Coral Gables
Conference on High Energy Physics, ed. Mintz, Perlmutter (Plenum Press, 1985);
L. Alvarez-Gaum\'e et al. Ref. [6]. 
\item
QCD corrections to this relation are given in S.P. Martin and M.T. Vaughn,
Phys. Lett. {\bf B318}, 331 (1993); D. Pierce and A. Papdopoulos, Nucl. Phys.
{\bf B430}, 278 (1994).
\item
T. Tsukamoto, K. Fujii, H. Murayama, M. Yamaguchi, and Y. Okada, Phys. Rev.
{\bf D51}, 3153 (1995).
\item
I. Hinchliffe, F.E. Paige, M.D. Shapiro, J. S\"oderqvist and W. Yao,
hep-ph/9610544.
\item
R. Arnowitt and P. Nath, Phys. Rev. Lett. {\bf 69}, 725 (1992); P. Nath and R.
Arnowitt, Phys. Lett. {\bf B289}, 368 (1992).
\item
J.L. Feng, M.E. Peskin, H. Murayama and X. Tata, Phys. Rev. {\bf D52}, 1418
(1995).
\item
R. Barbieri, S. Ferrara and C.A. Savoy, Phys. Lett. {\bf B119}, 343 (1982);
L. Hall, J. Lykken and S. Weinberg, Phys. Rev. {\bf D27}, 2359 (1983); P.
Nath, R. Arnowitt and A.H. Chamseddine, Nucl. Phys. {\bf B227}, 121 (1983); V.
Kaplunovsky and J. Louis, Phys. Lett. {\bf B306}, 269 (1993). 
\item 
N. Polonsky and A. Pomerol, Phys. Rev. {\bf D51}, 6532 (1995). 
\item
 H. Kim and C. Mu$\tilde n$os, hep-ph/9608214.
\item
A small correction of about 1\% in size proportional to $\delta_2 - \delta_1$
has been omitted from Eqs. (24) and (25).  See e.g. Kawamura et al., Ref. 25.
\item
M.M. Nojiri, Phys. Rev. {\bf D51}, 6281 (1995); M.E. Peskin, talk at YKIS95,
Kyoto (1995).
\item
J.F. Feng and D.E. Finnell, Phys. Rev. {\bf D49}, 2369 (1994). 
\item
H. Haber, Proc. of Beyond the Standard Model IV, ed. J. Gunion, T. Hans and J.
Ohnemus (World Scientific, Singapore, 1995).
\item
A. Bartl et al. hep-ph/9604221.
\item
M. Drees, Phys. Lett. {\bf B181}, 279 (1986); P. Nath and R. Arnowitt, Phys.
Rev. {\bf D39}, 2006 (1989); J.S. Hagelin and S. Kelley, Nucl. Phys. {\bf
B342}, 95 (1990).
\item
Y. Kawamura, H. Murayama and M. Yamaguchi, Phys. Lett. {\bf B324}, 52 (1994).
\item
A.E. Faraggi, Phys. Lett. {\bf B278}, 131 (1992); {\bf B302}, 202 (1993). 
\item
M. Dine, R. Leigh and A. Kagan, Phys. Rev. {\bf D48}, 4269 (1993).
\item
A. Font, L.E. Iba$\tilde n$ez, D. Lust and F. Quevedo, Phys. Lett. {\bf B245},
401 (1990); M. Cveti\u{c}, A. Font, L.E. Iba${\tilde n}$ez, D. Lust and F.
Quevedo, Nucl. Phys. {\bf B361}, 194 (1991); A. de la Macorra and G.G. Ross,
Nucl. Phys. {\bf B404}, 321 (1993); V. Kaplunovsky and J. Louis, Phys.
Lett. {\bf B306}, 269 (1993); R. Barbieri, J. Louis and M. Moretti, Phys.
Lett. {\bf B312}, 451 (1993); (Err. {\bf B316}, 632 (1993)); J.L. Lopez,
D.V. Nanopoulos and A. Zichichi, Phys. Lett. {\bf B319}, 451 (1993); S.
Ferrara, C. Kounnas and F. Zwirner, Nucl. Phys. {\bf B429}, 589 (1994)
(Err. {\bf B433}, 255 (1995)). 
\item
P. Candelas, M. Lynker and R. Schimmrigk, Nucl. Phys. {\bf B341}, 383 (1990);
J. Fuchs, A. Klemm, C. Scheich and M.G. Schmidt, Phys. Lett. {\bf B232}, 317
(1989).
\item
In Eq. (49) we have neglected the contribution in running the RGE from M$_G$
to M$_{str}$.  This contribution depends upon the particle content above
M$_G$ and on the Yukawa couplings, and hence requires fixing the Calabi-Yau
compactification to calculate it.  Thus our analysis here is meant to
illustrate what a LC could determine rather than being a detailed calculation
for a given model. (We also note that the choice of large ReT made
below moves M$_{str}$ closer to M$_G$ reducing this contribution).
\item
A. Klemm and S. Theisen, Nucl. Phys. {\bf B389}, 153 (1993); A. Font, Nucl.
Phys. {\bf B391}, 358 (1993); S. Hosono, A. Klemm, S. Theisen and S.-T. Yau,
Nucl. Phys. {\bf B433}, 501 (1995). 
\end{enumerate}
\end{document}